\documentclass[twocolumn,aps,pra,amsmath,amssymb,groupedaddress,superscriptaddress]{revtex4-1}

\usepackage{graphicx}
\usepackage{graphics}
\usepackage{dcolumn}
\usepackage{bm}
\usepackage{color}
\usepackage{soul}
\usepackage{textcomp}
\usepackage{soul}
\usepackage{rotating}
\usepackage{setspace}

\usepackage{microtype}
\usepackage[hyperindex=true,pdftitle={Molecular Excitation by Blackbody Radiation},
colorlinks=true,pagebackref=false,citecolor=blue,plainpages=false,pdfpagelabels,
linkcolor=blue,urlcolor=blue]{hyperref}
\def\di{{\rm d}}
\def\eu{{\rm e}}
\def\iu{{\rm i}}

\newcommand{\matrixel}[3]{\left< #1 \vphantom{#2#3} \right| #2 \left| #3 \vphantom{#1#2} \right>} 
\renewcommand{\v}[1]{\ensuremath{\mathbf{#1}}} 
\newcommand{\gv}[1]{\ensuremath{\mbox{\boldmath$ #1 $}}}  

\begin{document}

\title{
Classical Approach to Multichromophoric Resonance Energy Transfer
}
\author{Sebasti\'an Duque}
\affiliation{Grupo de F\'isica At\'omica y Molecular, Instituto de F\'{\i}sica,
Facultad de Ciencias Exactas y Naturales,
Universidad de Antioquia UdeA; Calle 70 No. 52-21, Medell\'in, Colombia.}
\affiliation{Chemical Physics Theory Group, Department of Chemistry and
Center for Quantum Information and Quantum Control,
\\ University of Toronto, Toronto, Canada M5S 3H6}

\author{Paul Brumer}
\affiliation{Chemical Physics Theory Group, Department of Chemistry and
Center for Quantum Information and Quantum Control,
\\ University of Toronto, Toronto, Canada M5S 3H6}
\author{Leonardo A. Pach\'on}
\affiliation{Grupo de F\'isica At\'omica y Molecular, Instituto de F\'{\i}sica,
Facultad de Ciencias Exactas y Naturales,
Universidad de Antioquia UdeA; Calle 70 No. 52-21, Medell\'in, Colombia.}

\begin{abstract}
A classical formulation of the quantum multichromophoric theory of resonance energy transfer
is developed on the basis of classical electrodynamics.
The  theory allows for the identification of a variety of processes of different order-in-the-interactions
that contribute to the energy transfer in molecular aggregates with intra-coupling
in donors and acceptor chromophores.
Enhanced rates in  multichromophoric resonance energy transfer are shown to be well
described by this theory.
Specifically, in a coupling configuration between $N_{\mathrm{A}}$ acceptors and $N_{\mathrm{D}}$
donors, the  theory correctly predicts an enhancement of the energy transfer rate dependent
on the total number of donor-acceptor pairs.
As an example, the theory, applied to the transfer rate in LH~II, gives results  in excellent
agreement with experiment.
Finally, it is explicitly shown that as long as linear response theory holds, the classical
multichromophoric theory formally coincides with the quantum formulation.
\end{abstract}

\date{\today}


\maketitle

\textit{Introduction}\textemdash Aspects of modern research on electronic resonant energy
transfer in photosynthetic light-harvesting systems have focussed on energy transfer as a
coherent collective phenomenon.
This feature has been highlighted as central to several transfer mechanisms, such as super
transfer \cite{LM10} and a network renormalization scheme \cite{RMS12}, and predicts dramatic
enhancements of  energy transfer rates \cite{KYR13}.
Qualitative arguments explaining such behavior often rely on
interactions within donors and acceptors that induce delocalization of the excitation and establish
quantum correlations, such as entanglement, between chromophores. As a consequence, this
observed unexpected rate enhancement has been widely attributed to quantum coherence of
acceptors and donors.

This purportedly quantum behavior at ambient conditions in photosynthetic light-harvesting systems
has contributed to the view that quantum effects play an important role in enhancing transport efficiency
in photosynthesis, and that these effects are somehow favored by evolutionary selection.
For example, arguments to explain transfer rate enhancements and irreversibility in light harvesting
complexes [such as  the Light Harvesting~II (LH~II)] as quantum processes involving superposition
and process coherence have
been proposed \cite{JNS07,JNS04,CS06, FD&14, OCF&08}, and the extent to which  enhancement is
quantum,  and is therefore incapable of being accounted for classically,  is  being extensively discussed
\cite{KYR13,PK&14, FD&14}.

In this letter we demonstrate that such enhanced rates are readily explained by a classical theory
that is reliant solely upon classical electrodynamics.
The resultant expressions retain the simplicity of F\"orster energy transfer formulae, while allowing
a straightforward interpretation of the origin of the enhanced energy transfer rates.
We apply this approach to calculate the energy transfer rate in both a model system and in LHII and
show that it  accurately describes enhanced multichromophoric energy transfer rates.
Since multichromophoric electronic energy transfer is also prevalent  in a large range of studies on
molecular systems such as DNA \cite{HX03} and proteins \cite{LS&03}, the theory is expected to be
useful in a wide variety of applications.

\textit{Quantum Multichromopric F\"orster's Resonance Energy Transfer}\textemdash
Note first the current quantum perspective on multichromophoric electronic energy transfer.
Consider the pairwise transfer of excitation from chromophore $\mathrm{D}$ to $\mathrm{A}$:
$
 \mathrm{D}^*+\mathrm{A}\rightarrow \mathrm{D}+\mathrm{A}^*,
$
where $\mathrm{D}^*$ ($\mathrm{D}$) is the excited (ground) state donor and $\mathrm{A}$ ($\mathrm{A}^*$)
is the ground (excited) state acceptor. From the single chromophoric F\"orster theory, the rate of energy
transfer from $\mathrm{D}$ to $\mathrm{A}$ is given by
$
 k_{\mathrm F}=\frac{J^2}{2\pi\hbar}\int_{-\infty}^{\infty} \di\omega\, E_\mathrm{D}(\omega) I_\mathrm{A}(\omega)
$
where $J$ is the electronic coupling between $\mathrm{D}$ and $\mathrm{A}$, $E_\mathrm{D}(\omega)$
is related to the normalized emission lineshape of the donor $\mathrm{D}$, and $I_\mathrm{A}(\omega)$
to the linear absorption cross section of the acceptor $\mathrm{A}$ \cite{J07}.

As long as the $\mathrm{D}$ and $\mathrm{A}$ molecules are well separated from one another,
inter-$\mathrm{D}$-$\mathrm{A}$ distances are larger than intra-$\mathrm{D}$ and intra-$\mathrm{A}$
distances, and well-defined $\mathrm{D}$ and $\mathrm{A}$ sites exists, so that the use of the rate
expression $k_{\mathrm F}$ is justified.
However, application of this single chromophoric theory to multichromophoric systems leads to errors
because transfer involves more than one pair of excitations, and because intra-$\mathrm{D}$ and
intra-$\mathrm{A}$ coherences that allow exciton delocalization over multiple chromophores are neglected.

These facts motivated a general quantum F\"orster-like rate expression for a set of
$\mathrm{D}_j$ ($j=1,\dots,N_\mathrm{D}$) donors and $\mathrm{A}_k$ ($k=1,\dots,
N_\mathrm{A}$) acceptors with intra-$\mathrm{D}$ and intra-$\mathrm{A}$  coherences, 
formulated in Ref.~\cite{JNS04}.
The expression can be cast as
\begin{align}
 \label{eq:2}
 k_\mathrm{F}^{\mathrm{MC}}=\sum_{j'j''}^{N_\mathrm{D}}\sum_{k'k''}^{N_\mathrm{A}}
 \frac{J_{j'k'}J_{j''k''}}{2\pi\hbar^2}\int_{-\infty}^\infty \di \omega\,
 E_\mathrm{D}^{j''j'}(\omega) I_\mathrm{A}^{k''k'}(\omega) \, ,
\end{align}
with $I_\mathrm{A}^{k''k'}(\omega)$ and $E_\mathrm{D}^{j''j'}(\omega)$ the absorption of acceptors
and the stimulated emission of donors, respectively.
The intra-$\mathrm{D}$ and intra-$\mathrm{A}$ coherences are said to be quantum,
arising from a  superposition of energy eigenstates, and to be responsible for the enhanced
transfer rate (e.g., Ref.~\cite{CS06}).
%

\textit{Classical Multichromophoric F\"orster's Resonance Energy Transfer}\textemdash
Classically, the donor is envisioned as an oscillating dipole of frequency
$\omega_\mathrm{D}$, and the acceptor as an absorber with oscillation frequency
$\omega_\mathrm{A}$.
The donor radiates an electric field that permeates the acceptor and the
acceptor absorbs energy from this field \cite{CPS75,NH06}.
Adopting this view, Kuhn \cite{Kuh70} and Silbey \textit{et al.} \cite{CPS75} derived, in the
1970's, F\"orster's transfer rate using a completely classical approach.
Specifically, they showed that the rate of energy transfer of a set of classically interacting
dipoles can be recast in a form identical to that of F\"orster theory \cite{CPS75, Kuh70}.
Here we significantly extend Refs. \cite{CPS75} and \cite{Kuh70} to obtain a  classical
description of multichromophoric energy transfer.

To do so, consider as above a set of
$N_\mathrm{D}$ donor molecules and $N_\mathrm{A}$ acceptor molecules, located at
$\mathbf{r}_{\mathrm{D}_j}$ and $\mathbf{r}_{\mathrm{A}_k}$, respectively.
The polarization of the $n^{th}$ molecule, at position $\mathbf{r}_n$, is proportional to the applied
field (linear response),
$
 \mathbf{p}_{n}(\omega)=
 \epsilon_0 \boldsymbol{\chi}_{n}(\omega) \mathbf{E}(\mathbf{r}_{n},\omega)\, ,
$
where $\mathbf{E}(\mathbf{r}_{n},\omega)$ is the
$\omega$ frequency component of the total electric field at $\mathbf{r}_n$ and $\boldsymbol{\chi}_n(\omega)$ is the
polarizability tensor of the $n^{th}$ molecule 
$(n=\mathrm{D}_1,\dots,\mathrm{D}_{N_\mathrm{D}},\mathrm{A}_1,\dots,\mathrm{A}_{N_\mathrm{A}})$.
The electric field at position $\mathbf{r}$ can be decomposed into an externally incident field
$\mathbf{E}^{\rm{ext}}$ and the sum of the fields produced by all others molecules in the aggregate.
In the non-radiative approximation, the electric field at point $\mathbf{r}$ due to the presence of a dipole
$\mathbf{p}$ at point $\mathbf{r}_0$ is
$ \mathbf{E}(\mathbf{r},\omega)=\frac{3\mathbf{\hat n}\mathbf{\hat n}-1}
 {4\pi\epsilon_0 |\mathbf{r}-\mathbf{r}_0|^3}\, \mathbf{p}(\omega)
 \equiv \Phi(\v{r}-\v{r}_0)\, \mathbf{p}(\omega)$ ,
where $\mathbf{\hat n}$ is the unit vector directed from
$\mathbf{r}_0$ to $\mathbf{r}$.
The polarization of each of the donor and acceptor molecules is
\begin{equation}
\label{eq:polarizationA}
\mathbf{p}_n(\omega)= \epsilon_0\boldsymbol{\chi}_{n}(\omega) \mathbf{E}^{\mathrm{ext}}(\mathbf{r}_n,\omega)
+ \epsilon_0\boldsymbol{\chi}_{n}(\omega) \sum_{n'} \Phi_{nn'}\, \mathbf{p}_{n'}(\omega),
\end{equation}
where $\Phi_{nn'}$ is the dipolar orientational coupling between molecules $n$ and $n'$
spanned by four blocks: the  $\Phi^\mathrm{D}_{jj'}$
block (denoted $\Phi^\mathrm{D}$ below) describes intra-D coupling between $\mathrm{D}_j$ and $\mathrm{D}_j'$
(for $j,j'=1,\dots,N_\mathrm{D}$), the block $\Phi^\mathrm{A}_{kk'}$ (denoted $\Phi^\mathrm{A}$ below) related to
intra-A coupling between  $\mathrm{A}_k$ and $\mathrm{A}_k'$
(for $k,k'=1,\dots,N_\mathrm{A}$) and the $\Phi_{jk}^\mathrm{DA}$ block (denoted $\Phi^\mathrm{DA}$) are the $\mathrm{D}_j$
and $\mathrm{A}_k$ interaction.
Here, the external field is only applied to the donors, so that
$\mathbf{E}^\mathrm{ext}(\mathbf{r}_n,\omega)=0$ for $n=\mathrm{A}_1,\dots,\mathrm{A}_{N_\mathrm{A}}$.
The case when the field impulsively excites all donors and acceptors
can be found in the Supplementary Material.

Although Eq.~\eqref{eq:polarizationA} is formulated in the frequency domain, it is clear that in the
time domain these processes are oscillatory (see below) and that the lifetime of the oscillations
depends upon the structure and values of $\boldsymbol{\chi}$.
For example, in a symmetric configuration in which the acceptors
have the same constant coupling $\Phi^\mathrm{A}_{kk'} = \phi^\mathrm{A}$, 
with identical acceptor response $\boldsymbol{\chi}_{\mathrm{A}_k}=\boldsymbol{\chi}_{\mathrm{A}}$,
the term related to the intra-A interactions in Eq.~(\ref{eq:polarizationA}) is
$
\epsilon_0 \boldsymbol{\chi}_{\mathrm{A}}
 (N_\mathrm{A}-1) \phi^\mathrm{A}\, {\mathbf{p}_\mathrm{A}}_{k}.
$
Despite the fact that this term already predicts an enhancement of the polarization
of the $k$-acceptor, it is shown below that this interaction need not be the one
responsible for the dramatic enhancement of the transfer rate.
Rather, it is the term in Eq.~(\ref{eq:polarizationA}) that allows every
acceptor to interact with every donor that is often significant (see Supplementary 
Material for further details).

Within classical electrodynamics, the Poynting vector
$\mathbf{S}=\mathbf{E}\times\mathbf{H}$ describes the energy flux density of the electromagnetic
field.
The rate of energy to or from a unit volume free of current or charges is
$ \dot{u}(\mathbf{r},t) = -\nabla\cdot\mathbf{S} $ and, using Maxwell's equations and
integrating over a volume enclosing the acceptor region, the rate of energy flow absorbed
by the acceptors is
\begin{equation}
\label{eq:QA}
 \dot Q(t)=\sum_{k=1}^{N_\mathrm{A}}
 \mathbf{E}(\mathbf{r}_{\mathrm{A}_k},t)\cdot \dot{ \mathbf{p}}_{\mathrm{A}_k}(t)\, ,
\end{equation}
and similarly for donors.
Here $\mathbf{p}_{\mathrm{A}_k}(t)$   denotes  the polarizability in the time domain
\cite{LL60,ZS10,VS&14} and  $\mathbf{E}(\mathbf{r}_{\mathrm{A}_k},t)$ labels
the total electric field at the position of the $k^{th}$ acceptor at time $t$.
$\dot Q(t)$ provides the time dynamics of energy transfer.
To see how it relates to F\"orster rate theory \cite{CPS75},
consider a set of $N_\mathrm{D}$ donors and $N_\mathrm{A}$ acceptors.
If each dipole is polarizable along a single axis, then 
$\mathbf{p}_n=p_n \mathbf{\hat n}_{n}$, $\boldsymbol{\chi}_{n}=\chi_{n}\mathbf{\hat n}_{n}\mathbf{\hat n}_n$
and if the external field is applied along this axis,
$\mathbf{E}_\mathrm{ext}(\mathbf{r}_n)=E_{n,\mathrm{ext}}\mathbf{\hat n}_{n}$,
then the polarization equation  \eqref{eq:polarizationA}, in the frequency domain,
can be conveniently expressed as $\mathsf{F}^{-1}\mathbf{p}=\mathbf{E}^{\rm{ext}}$,
where the polarizability matrix $\mathsf{F}^{-1}$ is defined as
$
 \mathsf{F}^{-1}_{nn'}=\left[\frac{\delta_{nn'}}{\epsilon_0\chi_n}-\Phi_{nn'} \right],
$
and   the polarization vector   is    $\mathbf{p}=[\mathbf{p}_\mathrm{D},\mathbf{p}_\mathrm{A}]^\mathrm{T}$  
with the scalar components $p_n(\omega)$, 
and the external applied field vector $\mathbf{E}^{\rm{ext}}$ has scalar components
$E_{n,\mathrm{ext}}$  (for
$n,n'=\mathrm{D}_1,\dots,\mathrm{D}_{N_\mathrm{D}},\mathrm{A}_1,\dots,\mathrm{A}_{N_\mathrm{A}}$).
The presence of off-diagonal elements $\mathsf{F}_{ij}$ implies that individual
chromophores cannot be excited independently.
Therefore, the excitation at one site spreads over other sites, which  can be
viewed as exciton delocalization within the classical picture.  

The rate of energy flow absorbed by the acceptors within this configuration is
$
\dot Q(t)= [\Phi^\mathrm{A}\mathbf{p}_\mathrm{A}(t)
+ \Phi^\mathrm{DA}\mathbf{p}_\mathrm{D}(t)] \cdot \dot{ \mathbf{p}}_{\mathrm{A}}(t)
$.
In order to compare with F\"orster's rate, $\dot{Q}(t)$ is transformed into the frequency 
domain, $\tilde{ \dot{Q}}(\omega)$, the oscillations in the transfer rate integrated out 
and the average value of the rate $\tilde{ \dot{Q}}(0)$ obtained \cite{ZS10}.  
Specifically, as shown in the Supplementary Material \cite{supplement} 
$
\tilde{ \dot{Q}}(0) = 2\epsilon_0 \,\mathrm{Im} \int_0^\infty  \mathrm{d}\omega\,
     \omega\, {\Phi^\mathrm{DA}}\mathbf{p}_\mathrm{D}^*(\omega)
     \cdot[ (\boldsymbol{\chi}_{\mathrm{A}}^{-1}(\omega)/\epsilon_0-
{\Phi}^\mathrm{A})^{-1} {\Phi^\mathrm{DA}} \mathbf{p}_\mathrm{D}(\omega)]
      ,
$
or, written explicitly
\begin{equation}
  \label{eq:classForster}
  \begin{split}
    \tilde{ \dot{Q}}(0) &=\sum_{jj'}\sum_{kk'} 2\epsilon_0 \Phi^\mathrm{DA}_{kj}\Phi^\mathrm{DA}_{k'j'} \int\limits_0^\infty
     \mathrm{d}\omega \,  I_\mathrm{A}^{kk'}(\omega) E_\mathrm{D}^{j j'}(\omega)
  \end{split}
\end{equation}
with $I_\mathrm{A}^{kk'}(\omega) = \omega\, \mathrm{Im}\, (\boldsymbol{\chi}_{\mathrm{A}}^{-1}(\omega)/\epsilon_0-
{\Phi}^\mathrm{A})^{-1}_{kk'}$
and $E_\mathrm{D}^{j j'}(\omega) = {p^*_\mathrm{D}}_j(\omega) {p_\mathrm{D}}_{j'}(\omega)$
related to the emission and absorption spectrum of the donors and acceptors.
This expression recovers the  form of the multichromophoric F\"orster expression \eqref{eq:2}. 
As in Eq.~\eqref{eq:2}, the intra-donor interaction in the F\"orster rate $k_{\mathrm{F}}$ is encoded in 
the definition of $I_\mathrm{A}^{kk'}(\omega)$ and $E_\mathrm{D}^{j j'}(\omega)$.
Additionally, if only a single donor and a single acceptor are present, Eq.~\eqref{eq:classForster} 
coincides with a single donor transferring energy to a single acceptor~\cite{ZS10}. 

\begin{figure}[h]
\includegraphics[width = \columnwidth]{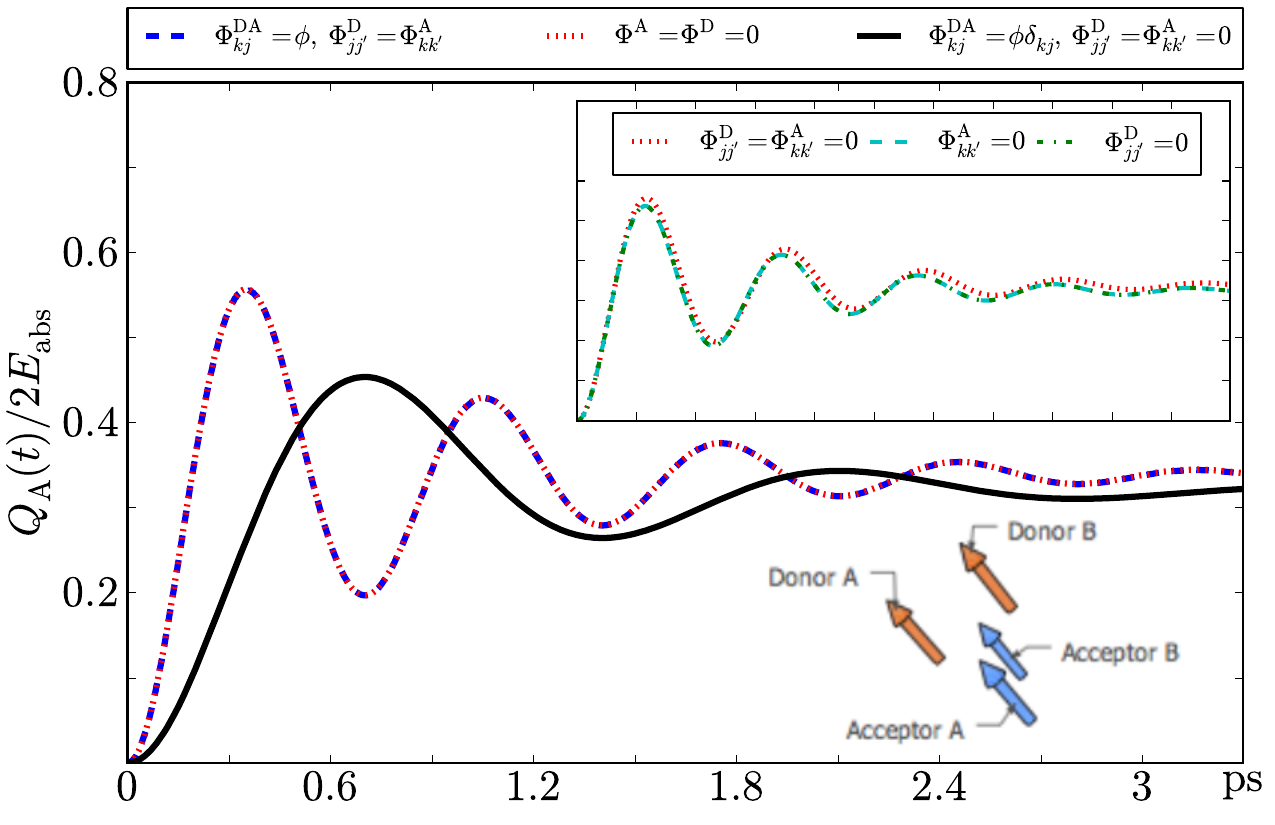}
\caption{
Normalized total energy in the acceptors for donors and acceptors in resonance at
$\omega_0=13317.2\, \rm{cm}^{-1}$ with dipole moments of $2.6 \,\rm{D}$.
All dipoles are separated $1 \,\rm{nm}$ from each
other.
The radiative decay rate is $\gamma=(0.8 \,\rm{ps})^{-1}$ for all dipoles.
Note that the blue dashed curve and dotted red curve lie atop one another.
Values of the parameters are typical for light-harvesting systems.
}
\label{fig:ClassFRET1}
\end{figure}

To show how classical electrodynamics gives the same transfer rate enhancement
as predicted by quantum arguments, consider a molecular aggregate model
comprised of two donors and two acceptors at the vertices of a tetrahedron, as shown
in the lower inset of Fig.~\ref{fig:ClassFRET1}.
The main panel of Fig.~\ref{fig:ClassFRET1} shows the normalized energy absorbed
by the acceptors [equation \eqref{eq:QA}, here denoted $Q_A(t)$]
with a single excited state and with Lorentzian lineshapes
$
 \chi(\omega) = 2(\epsilon_0\hbar)^{-1}
 \omega_0 |\gv{\mu}|^2/[(\omega_0-\omega
 + i\gamma/2)(\omega_0 +\omega +i \gamma/2)]
$,
where $\gv{\mu}$ is the transition dipole moment of the molecule, $\omega_0$ 
its transition frequency, $\gamma$ is radiative decay rate and 
$E_\mathrm{abs}=4\pi^2 \omega_0  |\gv{\mu}|^2N_\mathrm{D}/\hbar$ is the total 
energy absorbed by the donors from the electric field.
The donors are excited with a delta pulse in time.
Each molecule is polarized along a single polarization axis and all fields applied to the molecule are
along this axis of polarization.

The rate of energy transfer when the excitation is symmetrically delocalized over the interacting
dipoles, i.e., when the dipoles all interact
($\Phi^\mathrm{DA}_{kj}=\phi$ and
$\Phi_{jj'}^{\mathrm{D}}=\Phi_{kk'}^{\mathrm{A}}=\phi'$, $\phi$
and $\phi'$ constants) is shown as a dashed
blue line and is seen to be twice as fast as the case where the dipoles only communicate individually,
i.e., no donor and no acceptor interaction is present (the so-called ``direct transfer" case,
$\Phi^\mathrm{DA}_{kj} = \phi \delta_{kj}$, $\Phi_{jj'}^{\mathrm{D}} = \Phi_{kk'}^{\mathrm{A}} = 0$:
continuous black line).
Moreover, in the former fully connected case, not only is the frequency of the energy oscillation
(transfer rate)  faster but the amplitude of the energy oscillations is larger as well.
Thus, Fig.~\ref{fig:ClassFRET1} shows that classical electrodynamics predicts the same
enhancement of a factor of two as found in quantum approaches of excitonic energy transfer
\cite{LM10,KYR13}.

To understand the origin of this enhancement, we compare to the case
when there are no intra-interactions between donor or between acceptors, but where each
acceptor can interact with each donor ($\Phi^\mathrm{DA}_{jk} = \phi$, $\Phi_{jj'}^{\mathrm{D}} =
\Phi_{kk'}^{\mathrm{A}} = 0$: red dotted line).
The enhancement of the transfer rate is seen to be  virtually identical to  the case where
intra-interactions are allowed.
That is,
the enhancement here originates from the fact that all donors transfer to
all acceptors and not from the intra-interactions between acceptors or between donors,
an observation consistent with quantum results using the ``diagonal (secular) F\"orster rate"
model \cite{MAS99,JNS04,CC13}.

In the upper inset of Fig.~\ref{fig:ClassFRET1}, the case of vanishing intra-acceptor (or donor)
interactions in the presence of intra-donor (or acceptor) interactions is depicted by the dashed
cyan curve (or dot-dashed green curve).
 Although the effect here is small, it is clear that the transfer rate may indeed benefit from the
 lack of intra-donor or intra-acceptor interactions helping the energy transfer pathway.

\textit{Light Harvesting Complex II}\textemdash
To test the predictions of this classical theory, it is applied to calculate the transfer 
rate of LH II.
This complex is formed by 27 bacteriochlorophylls (BChls) arranged in two rings: eighteen of them
form the B850 ring with nine forming $\alpha\beta$-heterodimer subunits (here referred as the
acceptors), and the other nine the B800 ring (as the donor).
The LHII complex is described here by a set of interacting dipoles.
The couplings between the BChls in the B800 ring are much smaller than those in the B850 ring
\cite{JNS07,JC13} implying a monomeric structure for the B800 ring; hence the donor
is usually modelled as a single dipole \cite{CC13}.
The alternating transition dipole moment orientations within the B850 ring giving rise to the ninefold
symmetry is well depicted in  \cite{JS03}, as is the donor location.
Interdimer, intradimer coupling and site energies in the B850 ring are set as in \cite{CC13}.
The site energy of the two $\alpha\beta$-heterodimer subunits are
$E_{2n-1}=12406\,~ \mathrm{cm}^{-1}$ and $E_{2n}=12602\,~ \mathrm{cm}^{-1}$, the intradimer
coupling is $J_{2n-1,2n}=J_{2n,2n-1}=363 \,~ \mathrm{cm}^{-1}$ and the interdimer coupling
is $J_{2n+1,2n}=J_{2n,2n+1}=J_{1,18}=J_{18,1}=320 \,~ \mathrm{cm}^{-1}$ ($n=1,\dots,9$).
Intercomplex couplings between the elements comprising B850 are calculated using the point dipole approximation
with a transition dipole strength $\mu$ of $8.3\,\rm{D}$ and are related to the dipolar orientational
coupling $\Phi_{nn'}$ by $J_{nn'}=\mu^2 \Phi_{nn'}$. 

The environmental influence is included through the linear response function \cite{Ing02,GSI88}
$
 \chi_j(\omega) =
 2(\epsilon_0\hbar)^{-1}
 \omega_j \mu_j^2/ [-\omega^2-i\omega \tilde \gamma(i \omega)+\omega_j^2],
$
where $\mu_j$ is the transition dipole moment of molecule $j$, $\omega_j$ its transition
frequency and $\tilde \gamma(i \omega)$ is the Laplace transform of the damping kernel, related
to the spectral density    $\mathcal{J}(\omega)$  of the bath modes by
$
\tilde \gamma(i \omega) = \coth (\frac{1}{2}\hbar \omega \beta) \mathcal{J}(\omega).
$
For donor and acceptor molecules, independent identical baths are assumed and characterized by the
spectral density, $\hbar \mathcal{J}(\omega)=2\lambda\Lambda\omega/(\omega^2+\Lambda^2)$,
where $\lambda$ is the site reorganization energy of the donor (acceptor) and $\Lambda$ is
the inverse bath correlation time \cite{CC13}.
Setting $\lambda_\mathrm{D}=40\, \rm{cm}^{-1}$, $\lambda_\mathrm{A}=200~\rm{cm}^{-1}$,
$\Lambda=0.01\, \rm{fs}^{-1}$, and using a zero-mean-Gaussian-distributed energetic disorder of
$\sigma_\mathrm{D}=55\, \rm{cm}^{-1}$ in the donor and $\sigma _\mathrm{A}=290\, \rm{cm}^{-1}$
in the acceptors, reproduces the B800 and B850 absorption spectra at $T=300 \,\rm{K}$
\cite{JC13}.

\begin{figure}[t]
\includegraphics[width = \columnwidth]{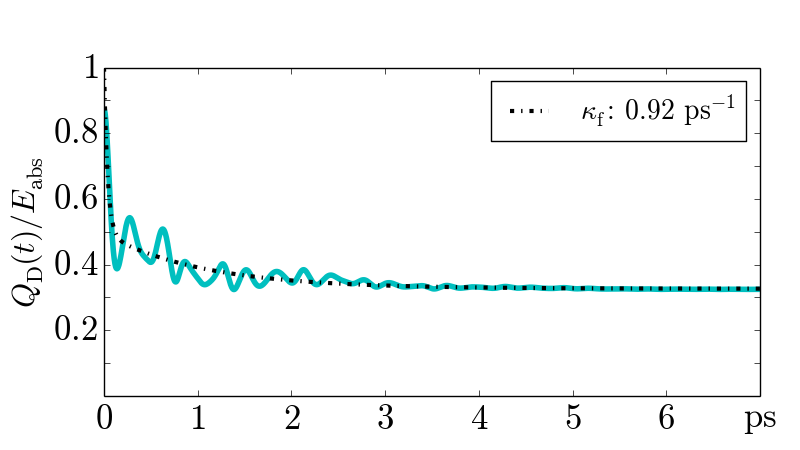}
\caption{
Normalized total energy within the B800 ring (donor) in LH2 initially excited with a delta pulse.
Dashed curve is a best fit to the oscillatory (cyan) result.
}
\label{fig:LH2}
\end{figure}
Figure \ref{fig:LH2} shows the normalized energy emitted by the donor [denoted $Q_\mathrm{D}(t)$] when 
excited initially with a delta pulse, for an ensemble of $10^4$ complexes.
The result of the simulation is best fit to a double exponential
decay $(a \mathrm{e}^{-\kappa_{\mathrm{f}} t}+b\mathrm{e}^{-\kappa_{\mathrm{uf}} t}+c)/(a+b+c)$
with an ultrafast component $\kappa_{\mathrm{uf}} = 22.36$~ps$^{-1}$, a fast component
$\kappa_{\mathrm{f}} = 0.92$~ps$^{-1}$, and the normalization
constants $a = 0.05$, $b = 0.12$ and $c = 0.1$.
The ultrafast component is associated with the sudden energy-absorption from the 
single-dipole donor to the acceptor ring. 
If the entire donor ring is included, the intra-donor dynamics modifies the transfer rates
and it is expected that the ultrafast component will be slower and the fast component's rate 
increases.
Work along this line is in progress.

The experimental transfer rate was reported to be
$\kappa_{\mathrm{exp}} \sim 1.25$~ps$^{-1}$ \cite{MCG97} while the quantum prediction, based
on a diagonal representation of multi-chromophoric energy transfer rate in Eq.~(\ref{eq:2}),
reported in Ref.~\cite{CC13}, is $\kappa_{\mathrm{qntm}} =0.7$~ps$^{-1}$.
Hence, the classical result obtained here of 0.92~ps$^{-1}$ predicts a transfer rate close to
the experimental rate, and is more accurate than the results predicted by the quantum 
calculation.
The fact that classical theory provides somewhat better results than does the quantum result
may arise from the fact that the classical transfer rate is obtained from the
time dynamics directly,  whereas the multichromophoric rate equation includes a number of
approximations (see Ref.~\cite{JC13} for details) and is calculated at $t\rightarrow \infty$.
Thus, the main dynamical features, such as the correct transfer rate, are not directly incorporated
into the quantum description.
This suggests that a full dynamic quantum calculation for the LH II, at the same level of the
classical one performed here, would be  of interest.
Furthermore, since it is shown above that Eqs.~\ref{eq:2} and \ref{eq:classForster} coincide, 
if no additional approximations are introduced, then both the quantum and classical results should
coincide.

The quantum-classical transition is discussed in the Supplementary Material. 
There it is shown that the two coincide when the assumption of linear response is valid and 
that the classical multichromophoric enhancement is still present in an effective single exciton 
regime considered by normalizing the energy in the entire aggregate. 

\textit{Comments }\textemdash
(a) To reconcile the above result with the supertransfer mechanism \cite{LM10}, note the standard 
quantum argument which proceeds as follows:   if coherence
is not present within the donor region, the incoherent Fermi-golden-rule rate of a donor
to transmit energy to the acceptor is
$\gamma_{\mathrm{D\rightarrow A}}\sim |\boldsymbol{\mu}_\mathrm{D}\cdot\boldsymbol{\mu}_\mathrm{A}|^2$.
Hence, for a pair of identical donors and a pair of identical acceptors
the total rate reads $\Gamma^{\mathrm{inc}}_{\mathrm{D\rightarrow A}}=2\gamma_{\mathrm{D\rightarrow A}}$.
However, if local coherence is present and the donor is in the symmetric ground state
$(\boldsymbol{\mu}_{1\mathrm{D}} + \boldsymbol{\mu}_{2\mathrm{D}} )/\sqrt{2}$ and
communicates with the corresponding state on the acceptor, the total rate
$
 \Gamma^{\mathrm{coh}}_{\mathrm{D\rightarrow A}} \sim \frac{1}{4} \bigg|
\boldsymbol{\mu}_{1\mathrm{D}} \cdot \boldsymbol{\mu}_{1\mathrm{A}} +
\boldsymbol{\mu}_{1\mathrm{D}} \cdot \boldsymbol{\mu}_{2\mathrm{A}} +
\boldsymbol{\mu}_{2\mathrm{D}} \cdot \boldsymbol{\mu}_{1\mathrm{A}} +
\boldsymbol{\mu}_{2\mathrm{D}} \cdot \boldsymbol{\mu}_{2\mathrm{A}}
 \bigg|^2,
$
so that $ \Gamma^{\mathrm{coh}}_{\mathrm{D\rightarrow A}} =
2\Gamma^{\mathrm{inc}}_{\mathrm{D\rightarrow A}}$.
Thus, the enhancement of the coherent rate $\Gamma^{\mathrm{coh}}_{\mathrm{D\rightarrow A}}$
comes from the terms $\boldsymbol{\mu}_{1\mathrm{D}} \cdot \boldsymbol{\mu}_{2\mathrm{A}}$
and $\boldsymbol{\mu}_{2\mathrm{D}} \cdot \boldsymbol{\mu}_{1\mathrm{A}}$, which
include the interactions between all donors and all acceptors.
Therefore, the enhancement that we obtained above, based on classical electrodynamics, is
precisely the one predicted by supertransfer \cite{LM10, KYR13} and corresponds to these 
terms in Eq.~(\ref{eq:polarizationA}).
Note that the classical theory formulated here also predicts additional processes that may
enhance or diminish energy transfer [see Supplementary Material].

(b) We note that the treatment in this letter has adopted a ``site basis" approach,
focusing on each dipole.
A generalized formulation could be used to study global donor or acceptor bright
or dark states, which would be obtained as eigenstates of the $\mathsf{F}$ matrix, and
used to define the initial conditions for the subsequent dynamical evolution.

In summary, a classical theory of multichromophoric electronic energy transfer was 
developed and shown it capable of producing the enhancement predicted by quantum-based 
approaches and that, as long as linear response holds, the classical 
approach coincides formally with the quantum description.  
Excellent results were also obtained for the LH~II case of one donor and multiple acceptors.
Further studies are  underway to display the utility of this approach in a variety
of other energy transfer scenarios.

\begin{acknowledgements}
The authors thank Professor Jianshu Cao, MIT, for providing data on LH~II, and Mr. Simon 
Axelrod and Dr. Aurelia Chenu for comments on an earlier version of this manuscript.   
This work was supported by NSERC Canada, by \textit{Comit\'e para el Desarrollo de la Investigaci\'on}
(CODI) of Universidad de Antioquia, Colombia under contract number E01651 and under
the \textit{Estrategia de Sostenibilidad 2014-2015} and by the \textit{Departamento Administrativo
de Ciencia, Tecnolog\'ia e Innovaci\'on} (COLCIENCIAS) of Colombia under the contract
number 111556934912.
\end{acknowledgements}

\bibliography{mcetprlv3}

\newpage

\begin{widetext}

\section*{Supplementary Material\\ Classical Approach to Multichromophoric Resonance Energy Transfer}

\section{I. Processes contributing to the Classical Energy Transfer Rate}

As in the main text, consider a set of $N_\mathrm{D}$ donor molecules and
$N_\mathrm{A}$ acceptor molecules, located at $\mathbf{r}_{\mathrm{D}_j}$
and $\mathbf{r}_{\mathrm{A}_k}$, respectively.
The polarization of each molecule is proportional to the applied
field (linear response)
$
 \mathbf{p}_{\mathrm{D}_j}(\omega)=
 \epsilon_0 \boldsymbol{\chi}_{\mathrm{D}_j}(\omega) \mathbf{E}(\mathbf{r}_{\mathrm{D}_j},\omega)\, ,
$
 $
 \mathbf{p}_{\mathrm{A}_k}(\omega)=
 \epsilon_0 \boldsymbol{\chi}_{\mathrm{A}_k}(\omega) \mathbf{E}(\mathbf{r}_{\mathrm{A}_k},\omega)\, ,
$
where $\mathbf{E}(\mathbf{r}_{\mathrm{D}_j},\omega)$ is the
$\omega$ frequency component of the total electric field at the position of the donor (similarly
for the acceptor) and $\boldsymbol{\chi}(\omega)$ is the polarizability tensor of the molecule.
The electric field at position $\mathbf{r}$ can be decomposed into an externally incident field
$\mathbf{E}^{\rm{ext}}$ and the sum of the fields produced by all others molecules in the aggregate.
In the non-radiative approximation, the electric field at point $\mathbf{r}$ due to the presence of a dipole
$\mathbf{p}_n$ at point $\mathbf{r}_n$ is
$ \mathbf{E}(\mathbf{r},\omega)=\frac{3\mathbf{\hat n}\mathbf{\hat n}-1}
 {4\pi\epsilon_0 |\mathbf{r}-\mathbf{r}_n|^3}\, \mathbf{p}_n(\omega)
 \equiv \Phi(\v{r}-\v{r}_n)\, \mathbf{p}_n(\omega)$ ,
where $\mathbf{\hat n}$ is the unit vector directed from $\mathbf{r}_n$ to $\mathbf{r}$. 
If the external field is zero in the region of the acceptors, the polarization of each of the donor and acceptor molecules is
\begin{equation}
 {\mathbf{p}_\mathrm{D}}_j(\omega)= \epsilon_0\boldsymbol{\chi}_{\mathrm{D}_j}(\omega)
 \mathbf{E}^{\rm{ext}}({\mathbf{r}_\mathrm{D}}_j,\omega)
 + \epsilon_0\boldsymbol{\chi}_{\mathrm{D}_j}(\omega)
 \sum_{j'\neq j}^{N_\mathrm{D}} \Phi^\mathrm{D}_{jj'}\, {\mathbf{p}_\mathrm{D}}_{j'}(\omega)
 + \epsilon_0 \boldsymbol{\chi}_{\mathrm{D}_j}(\omega)
  \sum_{k=1}^{N_\mathrm{A}} \Phi^\mathrm{DA}_{kj}\, {\mathbf{p}_\mathrm{A}}_k(\omega),
\end{equation}
and
\begin{equation}
  {\mathbf{p}_\mathrm{A}}_k(\omega)= \epsilon_0 \boldsymbol{\chi}_{\mathrm{A}_k}(\omega)
 \sum_{k'\neq k}^{N_\mathrm{A}} \Phi^\mathrm{A}_{kk'}\, {\mathbf{p}_\mathrm{A}}_{k'}(\omega)
 + \epsilon_0 \boldsymbol{\chi}_{\mathrm{A}_k}(\omega)
  \sum_{j=1}^{N_\mathrm{D}} \Phi^\mathrm{DA}_{kj}\, {\mathbf{p}_\mathrm{D}}_{j}(\omega),
\end{equation}
where $\Phi^\mathrm{DA}_{kj}$ is the dipolar coupling between $\mathrm{D}_j$ and $\mathrm{A}_k$, $\Phi^\mathrm{D}_{jj'}$
is the $\mathrm{D}_j$ and $\mathrm{D}_j'$ intra-D coupling and $\Phi^\mathrm{A}_{kk'}$ the $\mathrm{A}_k$ and
$\mathrm{A}_k'$ intra-A coupling.
Here $\mathbf{E}^{\rm{ext}}$ is the external field applied only to the donors.

To expose  the interplay between donors and acceptors, it is convenient to
subsitute the expression for $ {\mathbf{p}_\mathrm{D}}_j(\omega)$ into $ {\mathbf{p}_\mathrm{A}}_k(\omega)$,
giving
\begin{equation}
 \label{eq:polarizationA}
 \begin{split}
 {\mathbf{p}_\mathrm{A}}_k(\omega)&= \epsilon_0 \boldsymbol{\chi}_{\mathrm{A}_k}(\omega)
 \sum_{k'\neq k}^{N_\mathrm{A}} \Phi^\mathrm{A}_{kk'}\, {\mathbf{p}_\mathrm{A}}_{k'}(\omega)
 + \epsilon_0^2 \sum_{j=1}^{N_\mathrm{D}} \boldsymbol{\chi}_{\mathrm{A}_k}(\omega)\Phi^\mathrm{DA}_{kj}\,
 \boldsymbol{\chi}_{\mathrm{D}_j}(\omega) \mathbf{E}^{\rm{ext}}({\mathbf{r}_\mathrm{D}}_j,\omega)
 \\&+
 \epsilon_0^2
 \sum_{j=1}^{N_\mathrm{D}} \boldsymbol{\chi}_{\mathrm{A}_k}(\omega) \Phi^\mathrm{DA}_{kj}\,
 \boldsymbol{\chi}_{\mathrm{D}_j}(\omega)
 \sum_{j'\neq j}^{N_\mathrm{D}} \Phi^\mathrm{D}_{jj'}\, {\mathbf{p}_\mathrm{D}}_{j'}(\omega)
+
 \epsilon_0^2
 \sum_{j=1}^{N_\mathrm{D}} \boldsymbol{\chi}_{\mathrm{A}_k}(\omega) \Phi^\mathrm{DA}_{kj}\,
 \boldsymbol{\chi}_{\mathrm{D}_j}(\omega)
 \sum_{k'=1}^{N_\mathrm{A}} \Phi^\mathrm{DA}_{jk'}\, {\mathbf{p}_\mathrm{A}}_{k'}(\omega).
 \end{split}
\end{equation}
Further iterations are possible but Eq.~(\ref{eq:polarizationA}) already displays a number of
processes that enhance the polarizability at A$_k$, and hence can affect the energy transfer.
(i) The first term in Eq.~(\ref{eq:polarizationA}) will mediate the transfer of energy between A$_{k'}$ and A$_k$
via the interaction term $\Phi^\mathrm{A}_{kk'}$.
(ii) In the second term, the electric field $\mathbf{E}^{\rm{ext}}({\mathbf{r}_\mathrm{D}}_j) $ excites the donor
D$_j$ which can transfer part of the energy of the field to the acceptor A$_k$ via the interaction term 
$\Phi^\mathrm{DA}_{kj}$.
(iii) The third term describes how energy  in  donor D$_{j'}$ can flow into donor D$_{j}$ due to  the interaction term 
$\Phi^\mathrm{D}_{jj'}$,  and how part of this energy can transfer to  acceptor A$_k$ via  the interaction term $\Phi_{kj}$.
(iv) The last term describes  transfer of energy stored in  acceptor A$_{k'}$ to  donor $\mathrm{D}_j$,
assisted by the interaction $\Phi^\mathrm{DA}_{jk'}$, and the subsequent transfer from $\mathrm{D}_j$ 
to A$_{k}$ mediated by  $\Phi^\mathrm{DA}_{kj}$.
Processes (i) and (ii) are  first order in the interactions (via $\Phi^\mathrm{A}_{kk'}$ and $\Phi^\mathrm{DA}_{kj}$, 
respectively), while  (iii) and (iv) are  second order in the interactions (via $\Phi^\mathrm{DA}_{kj}\Phi^\mathrm{D}_{jj'}$ and
$\Phi^\mathrm{DA}_{kj}\Phi^\mathrm{DA}_{jk'}$, respectively).
If, in addition, the external field $\mathbf{E}^{\rm{ext}}$ is allowed to interact with the acceptors,
energy can flow directly into the acceptors; however, this situation not relevant for the present discussion.

\section{II. Explicit Derivation of the Classical Energy Transfer Rate}

Consider the rate of energy flow absorbed by the acceptors $\dot Q(t)$ given by
\begin{equation}
 \dot Q(t)=\sum_{k=1}^{N_\mathrm{A}}
 \mathbf{E}(\mathbf{r}_{\mathrm{A}_k},t)\cdot \dot{ \mathbf{p}}_{\mathrm{A}_k}(t)\, ,
\end{equation}
$\mathbf{p}_{\mathrm{A}_k}(t)$ denotes here the polarizability in the time domain
and  $\mathbf{E}(\mathbf{r}_{\mathrm{A}_k},t)$ labels the total electric field at the
position of the $k$-th acceptor at time $t$.
$\dot Q(t)$ provides the time dynamics of energy transfer.
If each dipole is polarizable along a single axis $\mathbf{\hat n}_{i}$ and the external field is applied along this axis,
then the acceptor polarization equation, in the frequency domain, can be written in term of the scalar
quantities  $\boldsymbol{\chi}_{i}=\chi_{i}\mathbf{\hat n}_{i}\mathbf{\hat n}_i$ and $\mathbf{p}_i=p_i \mathbf{\hat n}_{i}$ as
$
  {p_\mathrm{A}}_k(\omega)= \epsilon_0 \chi_{\mathrm{A}_k}(\omega)
 \sum_{k'\neq k}^{N_\mathrm{A}} \Phi^\mathrm{A}_{kk'}{p_\mathrm{A}}_{k'}(\omega)
 + \epsilon_0 \chi_{\mathrm{A}_k}(\omega)
  \sum_{j=1}^{N_\mathrm{D}} \Phi^\mathrm{DA}_{kj} {p_\mathrm{D}}_{j}(\omega)
$. 
Defining the vectors $\mathbf{p}_\mathrm{D}(\omega)$,
$\mathbf{p}_\mathrm{A}(\omega)$, $\boldsymbol{\chi}_{\mathrm{A}}$,
and $\mathbf{E}^\mathrm{ext}(\omega)$ with components $p_{\mathrm{D}_j }$,
$p_{\mathrm{A}_k }$, ${\chi_\mathrm{A}}_k$ and $E^\mathrm{ext}({\mathbf{r}_\mathrm{D}}_j,\omega)$ 
(with $j=1,\dots,N_\mathrm{D}$ and $k=1,\dots,N_\mathrm{A}$), respectively, and the matrices
$\Phi^\mathrm{A}$
and $\Phi$ with components $\Phi^\mathrm{A}_{kk'}$ and $\Phi^\mathrm{DA}_{kj}$, respectively,
the above equation can be rewritten in the compact form
$
  \mathbf{p}_\mathrm{A}(\omega) = \epsilon_0 \boldsymbol{\chi}_{\mathrm{A}}(\omega)[
 \Phi^\mathrm{A} \mathbf{p}_\mathrm{A}(\omega) + \Phi^\mathrm{DA} \mathbf{p}_\mathrm{D}(\omega)]
$.
 Hence,  the linear relationship of the acceptor polarization to the donor's is
$
\mathbf{p}_\mathrm{A}(\omega) = (\boldsymbol{\chi}_{\mathrm{A}}^{-1}/\epsilon_0-
\Phi^\mathrm{A})^{-1}\Phi^\mathrm{DA} \mathbf{p}_\mathrm{D}
$.

The rate of energy flow absorbed by the acceptors within this configuration is
$
 \dot Q(t)= [{\Phi}^\mathrm{A}\mathbf{p}_\mathrm{A}(t)
 +{\Phi^\mathrm{DA}}\mathbf{p}_\mathrm{D}(t)] \cdot \dot{ \mathbf{p}}_{\mathrm{A}}(t)
$.
In order to compare with F\"orster's rate, $\dot{Q}(t)$ is transformed into the frequency
domain,
\begin{equation}
\tilde{ \dot{Q}}(\omega) = - \iu \int_{-\infty}^{\infty} \di\omega' \,\omega' \, 
\left[ \Phi^\mathrm{A} \mathbf{p}_\mathrm{A}(\omega-\omega')
+\Phi^\mathrm{DA} \mathbf{p}_\mathrm{D}(\omega-\omega')\right] \cdot \mathbf{p}_\mathrm{A}(\omega').
\end{equation}

To compare with F\"orster rate, the oscillations need to be integrated out.
This is accomplished by taking the $\omega=0$ component $\tilde{ \dot{Q}}(0)$
and, using the fact that
$\mathbf{p}_i^*(\omega)=\mathbf{p}_i(-\omega)$ and 
 $
\mathbf{p}_\mathrm{A}(\omega) = 
(\boldsymbol{\chi}_{\mathrm{A}}^{-1}/\epsilon_0-{\Phi}^\mathrm{A})^{-1} {\Phi^\mathrm{DA}} \mathbf{p}_\mathrm{D}
$,
\begin{align}
\tilde{ \dot{Q}}(0) = -\iu \epsilon_0 \int_{-\infty}^\infty
     \mathrm{d}\omega \, \omega &\,\left[ {\Phi}^\mathrm{A}
       \mathbf{p}_\mathrm{A}^*(\omega)\right] \cdot
     \mathbf{p}_\mathrm{A}(\omega)\nonumber\\
     &-\iu \epsilon_0 \int_{-\infty}^\infty \mathrm{d}\omega\,
     \omega\, {\Phi^\mathrm{DA}}\mathbf{p}_\mathrm{D}^*(\omega)
     \cdot[ (\boldsymbol{\chi}_{\mathrm{A}}^{-1}(\omega')/\epsilon_0-
{\Phi}^\mathrm{A})^{-1} {\Phi^\mathrm{DA}} \mathbf{p}_\mathrm{D}(\omega')]
\end{align}
The first term is identically zero. After rearranging terms using the symmetry properties of the integral,
\begin{align}
\tilde{ \dot{Q}}(0) = 2\epsilon_0 \,\mathrm{Im} \int_0^\infty  \mathrm{d}\omega\,
     \omega\, {\Phi^\mathrm{DA}}\mathbf{p}_\mathrm{D}^*(\omega)
     \cdot[ (\boldsymbol{\chi}_{\mathrm{A}}^{-1}(\omega)/\epsilon_0-
{\Phi}^\mathrm{A})^{-1} {\Phi^\mathrm{DA}} \mathbf{p}_\mathrm{D}(\omega)].
\end{align}
Expanding the inner products
and defining
$I_\mathrm{A}^{kk'}(\omega) = \omega\,\mathrm{Im}\, (\boldsymbol{\chi}_{\mathrm{A}}^{-1}(\omega)/\epsilon_0-
{\Phi}^\mathrm{A})^{-1}_{kk'}$
and $E_\mathrm{D}^{j j'}(\omega) = {p^*_\mathrm{D}}_j(\omega) {p_\mathrm{D}}_{j'}(\omega)$, the above equation becomes
\begin{equation}
  \label{eq:classForster}
  \begin{split}
    \tilde{ \dot{Q}}(0) &=\sum_{jj'}\sum_{kk'} 2\epsilon_0 \Phi^\mathrm{DA}_{kj}\Phi^\mathrm{DA}_{k'j'} 
    \int\limits_0^\infty
     \mathrm{d}\omega \, I_\mathrm{A}^{kk'}(\omega) E_\mathrm{D}^{j j'}(\omega),
  \end{split}
\end{equation}
which corresponds to Eq.~(4) in the main text.

\section{III. Explicit Derivation of the Quantum Energy Transfer Rate}

Our classical approach is a generalization of the framework presented in Ref. [17],
and we follow that approach below to establish the quantum-classical connection, 
for multichromophoric electronic energy transfer,  within linear response theory.
Consider the interaction Hamiltonian
\begin{align}
  \label{eq:Hint}
  \hat{H}_\mathrm{int}=&\frac{1}{2}\sum_{j=1}^{N_\mathrm{D}}\sum_{k=1}^{N_\mathrm{A}}
  J_{jk} (\hat d_{\mathrm{D}_j} \hat d_{\mathrm{A}_k}+\mathrm{h.c.})
+\frac{1}{2}\sum_{j\neq j'} \Delta_{jj'}^\mathrm{D} \hat d_{\mathrm{D}_j} \hat d_{\mathrm{D}_{j'}}
  +
  \frac{1}{2}\sum_{k\neq k'} \Delta_{kk'}^\mathrm{A} \hat d_{\mathrm{A}_k} \hat d_{\mathrm{A}_{k'}}
-\sum_{j=1}^{N_\mathrm{D}} E_j(t)  \hat d_{\mathrm{D}_j},
\end{align}
where $\hat d_{\mathrm{D}_j,\mathrm{A}_k}$ are the dipole operators for the donor
$j$ (acceptor $k$), defined as $\hat d_{\mathrm{D}_j}=(|0\rangle\langle D_j|+\mathrm{h.c.})$ for donor state $|D_j\rangle$
(similarly for acceptors), $E_j(t)$ the external field acting on the donor $j$.  
The interaction Hamiltonian in \eqref{eq:Hint} coincides with the Hamiltonian used in quantum
MCFRET calculations when  working in the site basis.

Up to first order in perturbation theory, the time evolution of the polarization operators
$\hat d_{\mathrm{D}_j,\mathrm{A}_k}$ is well described by linear response theory.
Within this approach, the polarization of the donor
$j$ is
\begin{align}
  p_{\mathrm{D}_j} &= \int_{-\infty}^{\infty} \di t'\, \bigg [ R_{\mathrm{D}_j}(t,t')E_j(t')
  +   \sum_{jk}J_{jk}  R_{\mathrm{D}_{j}\mathrm{A}_k}(t,t')
  + \sum_{j'\neq j} \Delta_{jj'}^\mathrm{D} R_{\mathrm{D}_{j}\mathrm{D}_{j'}}(t,t') \bigg ],
\end{align}
where  the linear response functions are of the general form
\begin{align}
  \label{eq:1}
  R_\alpha(t,t')=-\frac{\iu}{\hbar} \lim_{\eta \to 0} f_\alpha(t,t')\eu^{\eta(t'-t)}\theta(t-t')
\end{align}
and the corresponding functions are
\begin{align}
  \label{eq:3}
  f_{\mathrm{D}_j}(t,t')&=\matrixel{\psi}{[\hat d_{\mathrm{D}_j}(t'),\hat d_{\mathrm{D}_j}(t)]}{\psi},
\\
  f_{\mathrm{D}_{j}\mathrm{A}_k}(t,t')&=\matrixel{\psi}{[\hat d_{\mathrm{D}_j}(t),\hat d_{\mathrm{D}_j}(t')]}{\psi}
  \matrixel{\psi}{\hat d_{\mathrm{A}_{k}}(t')}{\psi},
\\
  f_{\mathrm{D}_{j}\mathrm{D}_{j'}}(t,t')&=
  \matrixel{\psi}{[\hat d_{\mathrm{D}_j}(t),\hat d_{\mathrm{D}_j}(t')]}{\psi} \matrixel{\psi}{\hat d_{\mathrm{D}_{j'}}(t')}{\psi}.
\end{align}
As in Ref. [17], terms $\matrixel{\psi}{\hat d_{\mathrm{D}_{j'}}(t')}{\psi}$ and
$ \matrixel{\psi}{\hat d_{\mathrm{A}_{k}}(t')}{\psi} $ are replaced by
$p_{\mathrm{D}_{j'}}(t')$ and $p_{\mathrm{A}_{k}}(t')$, respectively, and, since
$\chi_{\mathrm{D}_j}(t,t')= \matrixel{\psi}{[\hat d_{\mathrm{D}_j}(t),\hat d_{\mathrm{D}_j}(t')]}{\psi}$,
the expression for the donor polarization \eqref{eq:Hint} in Fourier space becomes
\begin{align}
  \label{eq:6}
  \tilde p_{\mathrm{D}_j}(\omega)=\chi_{\mathrm{D}_j}(\omega)E(\omega)&+
  \sum_{k=1}^{N_\mathrm{A}} J_{jk} \chi_{\mathrm{D}_j}(\omega)  \tilde p_{\mathrm{A}_k}(\omega)
+
  \sum_{j'\neq j} \Delta^\mathrm{D}_{jj'} \chi_{\mathrm{D}_j}(\omega) \tilde p_{\mathrm{D}_j}(\omega).
\end{align}
This expression coincides with the classical equation for donor
polarization in our classical approach [c.f. Eq. (1) in this Supplementary Material], with 
the various matrices now explicitly defined.  
Using the same method, the expression for the acceptor $\tilde p_{\mathrm{A}_k}(\omega)$ i
s similarly found to coincide in the classical and quantum pictures. 

As we are interested in the energy absorbed by the acceptors as a function of time, by
applying linear response, we have
\begin{align}
  \label{eq:7}
  Q_\mathrm{A}(t)&=
  -\frac{\iu}{\hbar}\sum_{k=1}^{N_\mathrm{A}}
  \int_{-\infty}^{t}\di t'\, \matrixel{\psi}{[\hat{H}_{\mathrm{A}_k},\hat{H}_\mathrm{int}(t')]}{\psi}
\\
\begin{split}
  &=-\frac{\iu}{\hbar}\sum_{k=1}^{N_\mathrm{A}}
  \int_{-\infty}^{t}\di t'\, \bigg (\sum_{j'=1}^{N_\mathrm{D}}\sum_{k'=1}^{N_\mathrm{A}} J_{j'k'}
\matrixel{\psi}{[\hat{H}_{\mathrm{A}_k},\hat{d}_{\mathrm{D}_{j'}}(t')\hat{d}_{\mathrm{A}_{k'}}(t')]}{\psi}
\\
  & \hphantom{=}+ \sum_{k'\neq k''} \Delta_{k'k''}^\mathrm{A}
  \matrixel{\psi}{[\hat{H}_{\mathrm{A}_k},
  \hat{d}_{\mathrm{A}_{k'}}(t')]}{\psi} \matrixel{\psi}{\hat{d}_{\mathrm{A}_{k''}}(t')}{\psi} \bigg).
  \end{split}
\end{align}
Using  Ehrenfest's theorem, the expectation values can be related to the classical
polarization
$\dot p_{\mathrm{A}_k}=-\iu\hbar^{-1}\matrixel{\psi}{[\hat{H}_{\mathrm{A}_k},\hat d_{\mathrm{A}_k}]}{\psi}$, giving
\begin{align}
  \label{eq:9}
  Q_\mathrm{A}(t) =&\sum_{j=1}^{N_\mathrm{D}} \sum_{k=1}^{N_\mathrm{A}} J_{jk}
  \int_{-\infty}^{t}\di t'\, p_{\mathrm{D}_j}(t')  \dot p_{\mathrm{A}_k}(t')
  + \sum_{k\neq k'} \Delta_{kk'}^\mathrm{A}\int_{-\infty}^{t}\di t'\, p_{\mathrm{A}_{k'}}(t')  \dot p_{\mathrm{A}_k}(t')\, ,
\end{align}
or, equivalently,
\begin{align}
  \label{eq:10}
   \dot Q_\mathrm{A}(t) &=\sum_{j=1}^{N_\mathrm{D}} \sum_{k=1}^{N_\mathrm{A}} J_{jk} \,  p_{\mathrm{D}_j}(t)
   \dot p_{\mathrm{A}_k}(t)
  +
  \sum_{k\neq k'} \Delta_{kk'}^\mathrm{A}\, p_{\mathrm{A}_{k'}}(t)  \dot p_{\mathrm{A}_k}(t)\, .
\end{align}
Again, this expression coincides, for a set of dipoles, with that of the energy rate absorption in equation 
(3) of the main text,  successfully extending the relationship between classical and quantum treatments 
to the case of multichromophoric electronic energy transfer.

An auxiliary issue relates to how one can   guarantee the level of single exciton regime in the classical case.
Although our approach does not define the effective single exciton case, in this work interest is in the
normalized total energy absorbed by the acceptors
$Q_\mathrm{A}/E_\mathrm{abs}$, where $E_\mathrm{abs}$ is the total
energy absorbed by the donors from the electric field.
As $\mathbf{p}_\mathrm{D}$  increases with the incident electric field,
$E_\mathrm{abs}$ also increases.  Thus $Q_{\mathrm{A}}/E_\mathrm{abs}$ remains normalized.  

Note that $E_\mathrm{abs}$ certainly depends on the number of donors $N_\mathrm{D}$
and therefore, our results point out that the enhancement is
insensitive to this normalization. 
Moreover, the enhancement is in the rate of energy transfer, and not necessarily in the
amount of  energy that is being transferred (see Fig. 1 in the manuscript).   
Consider then Eq.~(\ref{eq:10}) of this section and, for example,  the rather artificial, highly symmetric
case where $N_\mathrm{D}=N_\mathrm{A}=N$, $p_{\mathrm{D}_j}(t)=p_{\mathrm{D}}(t)$ and
$p_{\mathrm{A}_j}(t)=p_{\mathrm{A}}(t)$ with $\Delta_{kk'}=0$. Then,
\begin{align}
   \dot Q^{\mathrm{mcp}}_\mathrm{A}(t) & = J N^2 p_{\mathrm{D}}(t) \dot p_{\mathrm{A}}(t)
   \quad \mathrm{for}\,J_{kj}=J
   \nonumber \\
   \dot Q^{\mathrm{dir}}_\mathrm{A}(t) & = J N p_{\mathrm{D}}(t) \dot p_{\mathrm{A}}(t)
   \quad \mathrm{for}\,J_{kj}=J\delta_{kj}
\end{align}
Here the $\mathrm{mcp}$ superscript denotes the multichromophoric result and $\mathrm{dir}$ 
the direct result.  
Note that even if $Q_\mathrm{A}(t)$ is normalized by either $N$ or $N^2$, as long the
normalization factor is the same for the symmetric multichromophoric case $J_{kj}=J$
and for the case of direct transfer $J\delta_{kj}$, the ratio
$\dot Q^{\mathrm{mcp}}_\mathrm{A}(t) /\dot Q^{\mathrm{dir}}_\mathrm{A}(t) = N$.
Thus, the quantum-mechanically-predicted enhancement is present also in the
classical case regardless of the normalization condition used  to mimic the single
exciton regime.

\section{IV. Quantum/Classical Energy Transfer Rate for General Initial Conditions}

As in Ref. [4], consider the multichromophoric situation of a set of
$\mathrm{D}_j$ ($j=1,\dots,N_\mathrm{D}$) donors and $\mathrm{A}_k$ ($k=1,\dots,
N_\mathrm{A}$) acceptors with a coupling Hamiltonian equation \eqref{eq:Hint}
without the external electric field, with the initial state set
by the general initial density operator $\hat{\rho}(0)=\mathcal{N}[\hat{\rho}_\mathrm{D}(0)
+\hat{\rho}_\mathrm{A}(0)]\hat{\rho}_\mathrm{D}^g\hat{\rho}_\mathrm{A}^g$. 
Here $\mathcal N$ is a normalization constant, 
$\hat{\rho}_\mathrm{D}(0)=\sum_{j,j'}|D_j\rangle\langle D_{j'}|$, and
$\hat{\rho}_\mathrm{A}(0)=\sum_{k,k'}|A_k\rangle\langle A_{k'}|$.
Note that intra-D and intra-A interactions are included in the interaction Hamiltonian.
Expanding $\hat{\rho}_\mathrm{A}(t)=\sum_{k=1}^{N_\mathrm{A}}\mathrm{Tr}_\mathrm{B}
\{\langle A_k| \hat{\rho}(t) |A_k\rangle\}$  to second order in $H_{\mathrm{int}}$, tracing 
over the identical local baths B, and calculating its time derivative gives the rate of energy 
transfer as
\begin{align}
k_{\mathrm{F}}^{\mathrm{MC}} = &\sum_{j'j''}^{N_\mathrm{D}}\sum_{k'k''}^{N_\mathrm{A}}
 \frac{J_{j'k'}J_{j''k''}}{2\pi\hbar^2}\bigg [
 \int_{-\infty}^\infty \di \omega\,  E_\mathrm{D}^{j''j'}(\omega) I_\mathrm{A}^{k''k'}(\omega)
 -\int_{-\infty}^\infty \di \omega\,  E_\mathrm{A}^{k''k'}(\omega) I_\mathrm{D}^{j''j'}(\omega) \bigg ]
 \nonumber\\
&+ \sum_{k'k''}^{N_\mathrm{A}} \frac{\Delta_{k''k'}}{2\pi}
\int_{-\infty}^\infty \di \omega\, L_\mathrm{A}^{k''k'}(\omega)
\end{align}
with
$
L_\mathrm{A}^{k''k'} = \int_{-\infty}^{\infty}\di t\; \eu^{-\iu \omega t}
\mathrm{Tr}_\mathrm{b_\mathrm{A}}\{\eu^{-\iu H_\mathrm{A}^g t/\hbar} \langle A_{k'}|
\eu^{\iu H_\mathrm{A}^e t/\hbar}\rho_\mathrm{A}(0)$
$
\eu^{-\iu H_\mathrm{A}^e t/\hbar}|A_{k''} \rangle \rho_\mathrm{A}^g \}.
$

The first two terms are the net F\"orster rate of the energy going from donors to acceptors,
 decreased by the energy returning from the acceptors to the donors. 
A careful analysis and manipulation of the double sum in the last term shows that it
vanishes.

As in the case described in the main text, the classical transfer rate agrees with the
quantum expression as well.

\end{widetext}

\end{document}